# Entropic origin of dielectric relaxation universalities in heterogeneous materials (polymers, glasses, aerogel catalysts)

F.Brouers[1,1], O.Sotolongo-Costa[2,2], A.Gonzalez[2], J. P. Pirard[*]

[1] Department of Physics, University of Liège, 4000, Belgium & Chair of Complex Systems H.Poincaré, University of Havana, Cuba

[2] Faculty of Physics, Cátedra de Sistemas Complejos H.Poincaré, Universidad de La Habana, Cuba

Abstract

We have derived a universal relaxation function for heterogeneous materials using the maximum entropy principle for nonextensive systems. The power law exponents of the relaxation function are simply related to a global fractal parameter $\alpha$ and for large time to the entropy nonextensivity parameter q. For intermediate times the relaxation follows a stretched exponential behavior. The asymptotic power law behaviors both in the time and the frequency domains coincide with those of the Weron generalized dielectric function derived in the stochastic theory from an extension of the Levy central limit theorem. These results are in full agreement with the Jonscher universality principle and find application in the characterization of the dielectric properties of aerogels catalytic supports as well as in the problem of the relation between morphology and dielectric properties of polymer composites.

## 1 Introduction

These last decades, growth, aggregation and fragmentation models have been proposed and discussed with the aim of representing the (multi-)fractal geometry and the resulting scaling properties of a great variety of complex materials such as glasses, polymers, colloids, gels, self-similar porous and cellular materials [1]These studies cover a large range of physical (electrical, dielectric, optical, magnetic and mechanical) properties [2]. It has also been observed that universality is the major feature resulting from their scaling behavior [3],[4]. In the same context, it is more and more accepted that the Boltzmann statistics is not adequate to describe the macroscopic thermodynamic properties of natural phenomena when the effective microscopic interactions and the microscopic memory are long ranged due to complex non-equilibrium growth, aggregation or fragmentation processes. For that reason a generalized form of a nonextensive entropy $S_q$ known as the " Tsallis entropy" (T.E.) [5] has been used with some success although the fundamental basis of such a formulation is still the object of many discussions. [6],[7],[8].

Relaxation (dielectric, mechanical, magnetic...) in systems such as the ones we are considering here is a complex stochastic mechanism which depends both of the cluster geometric structure and the collective nature of the interactions. In most cluster models, relaxation is viewed as a complex hierarchical or stochastic process [9], [10]. Indeed in that case, an universal pattern independent of materials and microscopic model has been noticed for a long time [3],[4] and widely discussed in the literature. Experimentally one observes both for large $t$ and small $t$ a fractional power law dielectric response in the time domain:

---

[1] address: Department of Physics, University of Liège, 4000, Belgium & Chair of Complex Systems H.Poincaré, University of Havana, Cuba

[2] Faculty of Physics, Cátedra de Sistemas Complejos H.Poincaré, Universidad de La Habana, Cuba

[*] Department of Applied Chemistry, Liege University, Belgium

$$f(t) = -\frac{d\phi}{dt} = \begin{cases} (\omega_p t)^{-n} & \text{for } \omega_p t \ll 1 \quad (a) \\ (\omega_p t)^{-m-1} & \text{for } \omega_p t \gg 1 \quad (b) \end{cases} \qquad (1)$$

with $0 < n, m < 1$. Quite generally it is observed that $1 - n < m$. The loss peak frequency $\omega_p$ (frequency corresponding to the maximum of the dielectric plot) is material dependent. For intermediate $t$, the relaxation function $\phi(t)$ is usually fitted to a stretched exponential, the so-called Williams-Watts form: $\phi(t) \approx \exp(-(\omega_p t)^\alpha)$ which does not exhibit the algebraic large $t$ behavior (1b). The small $t$ (large $\omega$) exponent $n$ is related to small clusters relaxation and the large $t$ (small frequency) exponent $m-1$ describes the relaxation of large clusters. As it will appear in this work, they are not independent. This non-Debye behavior has its counterpart in the frequency domain where generalizations of the Cole and Cole dielectric functions with exponents linked to $n$ and $m$ are commonly used by experimentalists.

## 2.-Direct derivation of the universal dielectric function

The exact form of the universal dielectric function can be derived directly without using the traditional average of the Debye exponential relation over relaxation time [12]. It has been known for a long time that the stretched exponential distribution (Weibull distribution with $\alpha < 1$) which has been used under the name of Williams-Walls formula to fit dielectric relaxation can be obtain by minimizing the B.E. $S_{q=1}$ subject to two constraints on $x^\alpha$ and $\ln x$. We have shown that by generalizing this procedure using the T.E $S_q$, the $q$-constraints on $x^\alpha$ and $\ln_q(x)$ and the properties of $\ln_q(x)$ and $\exp_q(x)$, we obtain easily the density and distribution function and its density generalizing the Weibull distribution and density:

$$F_{\alpha,q}(x) = 1 - \left[1 + \xi \frac{x^\alpha}{<x^\alpha>_q}\right]^{-\frac{1}{\xi}}, \qquad f_{\alpha,q}(x) = \frac{\alpha x^{\alpha-1}}{<x^\alpha>_q}\left[1 + \xi \frac{x^\alpha}{<x^\alpha>_q}\right]^{-\frac{1}{\xi}-1} \qquad (2)$$

This distribution belongs to the domain of attraction of the one-sided completely asymmetric Levy stable law since with the heavy tail index

$$\mu = \alpha \xi^{-1} = \alpha \frac{2-q}{q-1} < 1 \qquad (3)$$

In the limit $q \to 1$ $(\xi \to 0)$, we recover the stretched exponential. These results can be used to represent non-Debye relaxation if we identify the random variable $X$ with the macroscopic waiting time $\Theta$ as defined by Weron et.al.[10]. We have :

$$F_\Theta(t) = \Pr(\theta < t) = \int_0^t f_{\alpha,q}(s)ds = 1 - \left[1 + \xi \frac{t^\alpha}{<\theta^\alpha>}\right]^{-\frac{1}{\xi}} \qquad (4)$$

which is a generalized Pareto law (Burr$_{XII}$: $B_{b,c}(x) = 1 - (1+x^b)^{-c}$, $x > 0$) [11]. The relaxation function $\phi(t)$ can be written as the survival probability of the non equilibrium initial state of the relaxing system. Its value is determined by the probability that the system as a whole will not make transition out of its original state until time $t$:

$$\phi_{\alpha,q}(t) = 1 - \Pr(\theta < t) = \Pr(\theta > t) = \left[1 + \xi(At)^{\alpha}\right]^{-\frac{1}{\xi}} \tag{5}$$

with

$$A \equiv \omega_p = <\theta^{\alpha}>_q^{-1/\alpha} \tag{6}$$

We can write the response function

$$f(t) = -\frac{d\phi(t)}{dt} = \alpha A (At)^{\alpha-1} \left[1 + \xi(At)^{\alpha}\right]^{-1-\frac{1}{\xi}} = \frac{\alpha (At)^{\alpha}}{t} \phi^{\xi+1} \tag{7}$$

which gives the two power-law asymptotic behavior (1) observed in dipolar relaxation [3]: In the frequency range the susceptibility $\chi(\omega) = \chi'(\omega) + i\chi''(\omega)$ [12] obeys the Jonscher universal laws [3]

$$\lim_{\omega \to \infty} \frac{\chi''(\omega)}{\chi'(\omega)} = \cot(n\pi/2) \quad \lim_{\omega \to 0} \frac{\chi''(\omega)}{\chi'(0) - \chi'(\omega)} = \tan(m\pi/2) \tag{8}$$

with $n = 1 - \alpha$, m=$\alpha/\xi$ and $\xi = \dfrac{q-1}{2-q}$ The $h$-moment of the Burr$_{XII}$ distribution diverges if m=$\alpha/\xi < h$. As a consequence, the expectation value of the random variable $\Theta$ diverges if $m < 1$ or $q > \dfrac{2\alpha+1}{1+\alpha}$. In the case of dipolar systems, most of the empirical data for the power-law exponents $n$ and $m$ are in the range [0,1] and can be accounted for using the heavy-tailed Burr$_{XII}$ waiting-time distribution with $\alpha/\xi \leq 1$ for which $<\theta> = \infty$. The scale parameter $A$ which is related to the the peak frequency $\omega_p$ of the response function in the frequency domain is materials and $q$-dependent and is simply related to the escort probability average of the random quantity $\theta^{\alpha}$, becomes the natural finite time scale in a physical system (for example the dipolar relaxation) which "chooses" [10] the regime where the usual waiting time average is infinite In other words when $\alpha/\xi \leq 1$, a scale can be defined only by using the escort average procedure and defining a $q$-dependent time scale as $A^{-1} = <\theta^{\alpha}>_q^{1/\alpha}$.

### 3.- Conclusion

Our model based on extremization of the nonextensive T.E. gives a physical interpretation of the mathematical parameters of the Weron stochastic theory [10],[13] and opens new paths to understand the ubiquity of self-similarity and power laws in the relaxation of large classes of materials in terms of their fractal and nonextensive properties. The universal Burr$_{XII}$ relaxation function can fit the relaxation and dielectric behavior of a large number of data reported in Jonsher [4]. Most importantly a finite scale can be defined as a $q$-expectation value using the concept of escort probability.This situation is typical of "frustrated" systems like glasses, polymers porous materials, heterogeneous surfaces, gels, etc. More details can be found in [8],[10],[12].